\documentclass[pra,superscriptaddress,twocolumn,showpacs,amsmath,aps]{revtex4}
\usepackage{graphicx,color}
\usepackage{bm}
\usepackage[hypertex]{hyperref}

\newcommand{\be}{\begin{equation}}
\newcommand{\ee}{\end{equation}}
\newcommand{\bea}{\begin{eqnarray}}
\newcommand{\eea}{\end{eqnarray}}
\newcommand{\bsube}{\begin{subequations}}
\newcommand{\esube}{\end{subequations}}

\begin{document}

\title{ Quantum transfer through a non-Markovian environment under frequent measurements and Zeno effect}
\author{Luting Xu}
\affiliation{Department of Physics, Beijing Normal University,
Beijing 100875, China}
\author{Yunshan Cao}
\affiliation{Department of Physics, Beijing Normal University,
Beijing 100875, China}
\author{Xin-Qi Li}
\email{lixinqi@bnu.edu.cn}
\affiliation{Department of Physics, Beijing Normal University,
Beijing 100875, China}
\author{YiJing Yan}
\affiliation{Department of Chemistry, Hong Kong University of
Science and Technology, Kowloon, Hong Kong}
\author{Shmuel Gurvitz}
\email{shmuel.gurvitz@weizmann.ac.il}
\affiliation{Department of Particle Physics and Astrophysics,
Weizmann Institute of Science, Rehovot 76100, Israel}
\affiliation{ Beijing Computational Science Research Center,
Beijing 100084, China}

\date{\today}

\begin{abstract}
We study transitions of a particle between two wells, separated by a reservoir, under the condition that the particle is not detected in the reservoir. Conventional quantum trajectory theory predicts that such no-result continuous measurement would not affect these transitions. We demonstrate that it holds only for Markovian reservoirs (infinite bandwidth $\Lambda$). In the case of finite $\Lambda$, the probability of the particle's interwell transition is a function of the ratio $\Lambda/\nu$, where $\nu$ is the frequency of measurements. This scaling tells us that in the limit  $\nu\to\infty$, the measurement freezes the initial state (the quantum Zeno effect), whereas for $\Lambda\to\infty$ it does not affect the particle's transition across the reservoir. The scaling is proved analytically by deriving a simple formula, which displays two regimes, with the Zeno effect and without the Zeno effect. It also supports a simple explanation of the Zeno effect entirely in terms of the energy-time uncertainty relation, with no explicit use of the projection postulate. Experimental tests of our predictions are discussed.
\end{abstract}

\pacs{03.65.Ta,03.65.Xp,73.63.$-$b,73.40.Gk}
\maketitle

It is well known that the unitary evolution of a quantum system is interrupted by measurement, so  the subsequent evolution of a system depends on the measurement record. Frequent measurements with intervals $\Delta t$ are of special interest. In the limit  $\Delta t\to 0$, they freeze the particle's motion (the quantum Zeno effect). This result is a consequence of the projection postulate applied to sequential measurements.

The Zeno effect looks very surprising since it reveals the dynamical impact of the projection postulate on quantum motion. Instead, one can try to attribute the Zeno effect to the influence of the measurement devices. At first it seems as though this cannot be the case. Indeed, due to the interaction with detectors, the system acquires the energy $\sim\hbar/\Delta t$, according to the energy-time uncertainty relation. As such, it is natural to expect an acceleration of the particle, instead of its freezing: the anti-Zeno effect \cite{Kur00}. Nevertheless, as demonstrated in this paper, the Zeno  effect can be entirely attributed to the energy-time uncertainty relation, without the explicit use of the projection postulate. It would make the Zeno effect much less surprising and in fact quite expectable.

The concept of continuous measurement is inherent in the quantum trajectory (QT) approach (informational evolution), which treats  quantum motion based on the results of intermediate measurements \cite{WM10}. It is therefore natural to investigate the Zeno-effect dynamics in this framework. A pronounced example of the informational evolution, was proposed for a two-state system (qubit), coupled to a continuously monitored reservoir under the condition that no signal is registered there \cite{Dali92,Kor06}. It was predicted that the qubit can change its state despite the null-result measurements. This was confirmed  in experiment with a superconducting phase qubit measured via tunneling \cite{Katz06}.

In this paper we study a different arrangement, with two distant localized states, connected by a common reservoir under continuous null-result monitoring \cite{SG11a,SG11c}. Predictions based on the QT approach for this case are even more dramatic: The system can display the transition between these two localized states via the reservoir, although the latter is under continuous null-result monitoring \cite{SG11c}. This result is highly counterintuitive and is a clear contradiction of the Zeno effect.
We therefore perform here a detailed analysis of these undetectable transitions, particularly in relation to the QT approach and  the Zeno effect.

\begin{figure}[h]
\center
\includegraphics[scale=0.4]{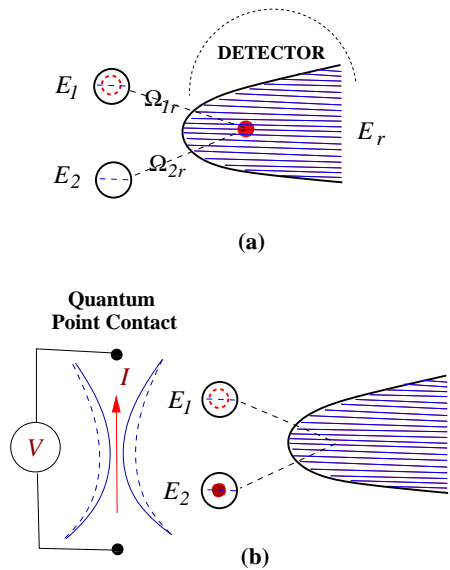}
\caption{(Color online) Two quantum dots coupled to the reservoir:
(a) The reservoir is monitored by a detector and (b) the quantum dots are
monitored by a point contact. The detector current $I$ increases when the electron leaves the dots.}
\label{fig1}
\end{figure}

Consider two quantum dots coupled to a reservoir,
monitored by an external detector, as shown in Fig.~\ref{fig1}(a)
or, equivalently, Fig.~\ref{fig1}(b), where
the point-contact (PC) current $I$ increases when the electron tunnels to the reservoir. In order to make the two setups fully equivalent, the PC detector should be placed symmetrically with respect to the dots.
The system is described by Hamiltonian \cite{SG11a,SG11c}
\begin{align}
H=\sum_{k=1,2,r}E_k|k\rangle\langle k|
+\sum_{j,r} \left[\Omega_{jr}|j\rangle\langle r|
     + {\rm H.c.} \right] \, ,  \label{a1}
\end{align}
where $j=1,2$ and $\Omega_{jr}$ is tunneling coupling of the dot $j$ with the reservoir. The states in the dots $|1(2)\rangle$ are localized and the reservoir states $|r\rangle$ are extended.

By diagonalizing the Hamiltonian (\ref{a1}) we find that all eigenstates are extended. However, if $E_1=E_2$ and
$\Omega_{1r}/\Omega_{2r}=\gamma =$ const [\onlinecite{note1}],
there exists one localized eigenstate of the Hamiltonian:
$|1'\rangle =\cos\beta\, |1\rangle -\sin\beta \,|2\rangle$,
where $\cos\beta =1/\sqrt{1+\gamma^2}$.
Indeed, $H|1'\rangle =E_1|1'\rangle$.
Note that the orthogonal state
$|2'\rangle =\sin\beta\,|1\rangle + \cos\beta\,|2\rangle$ is not an eigenstate of $H$, but decays to the reservoir.
Consider an initial state
$|\Psi (0)\rangle =\alpha_1|1'\rangle+\alpha_2|2'\rangle$.
We find that the probability of finding the electron in each of the dots,
conditioned on no electron in the reservoir,
is given by $\bar P_{1(2)}=(1+\gamma^2)^{-1}$.
Thus the electron, initially localized in the upper dot of Fig.~\ref{fig1}, can be found in the lower dot at $t\to\infty$ with non zero probability \cite{note3}. This implies a possibility of the inter-dot electron transitions through the reservoir, similar to transport through dark state in quantum optics (see the discussion in \cite{SG11c}).

Let us investigate how continuous monitoring of the reservoir affects these transitions. Consider again the electron in the linear superposition of the states $|1'(2')\rangle$. After a short time $\tau$ the wave function becomes
\begin{align}
|\Psi (\tau)\rangle =
\alpha_1^{}|1'\rangle
 +\alpha_2^{}\big[1-iH\tau-H^2\tau^2/2
 +\cdots\big]|2'\rangle
\label{a10}
\end{align}
where we expanded the evolution operator $\exp (-iH\tau)$ up to the second order in $\tau$. (From now on we adopt the units with $\hbar=1$.) The null-result measurement in the reservoir implies that the electron's wave function is projected on the two-dot subspace $|\Psi (\tau)\rangle\to \hat Q\,|\Psi (\tau)\rangle$, where $\hat Q=\big(|1'\rangle\langle 1'|+|2'\rangle\langle 2'|\big)/N$ and $N$ is a normalization factor. Therefore
\begin{align}
|\Psi_1\rangle =\hat Q\,|\Psi (\tau)\rangle
=\Big[\alpha_1^{}|1'\rangle+\alpha_2^{}\big(1
-C\tau^2\big)|2'\rangle\Big]/N_1 \,,
\label{a11p}
\end{align}
where $C={1\over2}\sum_r\big(\Omega_{1r}^2+\Omega_{2r}^2\big)$ and $N_1^2=1-2\,\alpha_2^2\,C\tau^2$.

After $n$ subsequent null-result measurements during time $t$, with $n=t/\tau$, we find
\begin{align}
|\Psi_n\rangle =\Big[\alpha_1^{}|1'\rangle
+\alpha_2^{}\big(1-C\tau^2\big)^n|2'\rangle \Big]/N_n \,,
\label{a13}
\end{align}
where $N_n^{}=\sqrt{1-2\, n\, \alpha_2^2 C\,\tau^2}$. Thus, in the limit $\tau\to 0$ and for $t$=const, one obtains
\begin{align}
|\Psi_n\rangle \to \alpha_1^{}|1'\rangle
+\alpha_2^{}|2'\rangle \equiv |\Psi (0)\rangle\, .
\label{a14}
\end{align}
This means that the continuous null-result monitoring of the reservoir [Fig.~\ref{fig1}(a)] or its indirect monitoring [Fig.~\ref{fig1}(b)] reveals the Zeno effect by preventing the electron's interdot transitions.

On the other hand, the conditional electron's dynamics
can be studied by the QT method, designed for such a type of problem.
Surprisingly, one arrives at the opposite conclusion \cite{SG11c}:
The continuous null-result monitoring of the reservoir does not
prevent the electron's interdot transitions.
In order to understand the disagreement between  predictions of
the QT approach and the Zeno effect, we present below a detailed
quantum-mechanical analysis of the continuous null-result
measurements for the setup in Fig.~\ref{fig1}.

Consider the electron wave function, written as
\begin{equation}\label{WF-1}
|\Psi(t)\rangle=b_1(t)|1\rangle+b_2(t)|2\rangle
+\sum_rb_r(t)|r\rangle,
\end{equation}
where $b_{1,2,r}(t)$ are the probability amplitudes of finding the electron in the dots and reservoir. Substituting Eq.~(\ref{WF-1}) into the Schr\"odinger equation $i\partial_t |\Psi(t)\rangle =H|\Psi(t)\rangle$ and performing the Laplace transform $\tilde b(\omega ) =\int_0^\infty b(t)\exp (i\omega t)dt$, we obtain the following system of algebraic equations for $\tilde b(\omega )$:
\begin{subequations}
\label{eqs}
\begin{align}
&(\omega-E_j)\tilde b_j(\omega)-\sum_r\Omega_{jr}\tilde b_r(\omega)=ib_j(0) \,,
\label{eqs1}\\
&(\omega-E_r)\tilde b_r(\omega)-\Omega_{1r}\tilde b_1(\omega)-\Omega_{2r}\tilde b_2(\omega)=0 \,,
\label{eqs2}
\end{align}
\end{subequations}
where $j=1,2$. (Similar equations can be written for multilevel systems \cite{ariel}.) The right-hand sides of these equations reflect the initial conditions, corresponding to the electron localized in the dots. Substituting $\tilde{b}_r(\omega)$ from Eq.~(\ref{eqs2}) into Eq.~(\ref{eqs1}) and replacing $\sum_r\to\int\varrho (E_r)dE_r$, where $\varrho(E_r)$ is the density of states, we obtain
\begin{align}
&(\omega-E_{j})\tilde b_j(\omega )-\sum_{j'}{\cal F}_{jj'}\tilde b_j(\omega )=i\,b_j(0)\, ,
\label{a2}
\end{align}
where
\begin{equation}  \label{eq:6}
\mathcal{F}_{jj'}(\omega)=
\int\frac{\Omega_{jr}\Omega_{j'r}}{\omega-E_r}\varrho(E_r)dE_r  \,.
\end{equation}

In many calculations the density of reservoir states is taken to be energy independent (the so-called wideband limit or Markovian reservoir).
Here we consider a finite-band spectrum by taking $\varrho(E_r)$ in the Lorentzian form,
\begin{align}
\varrho (E_r)=\varrho_0\Lambda^2/( E_r^2+\Lambda^2)\, ,
\label{lor}
\end{align}
while the coupling amplitudes are energy independent
$\Omega_{jr}=\Omega_j$. Then we obtain
\begin{align}
\mathcal{F}_{jj'}(\omega )
={\Lambda\sqrt{\Gamma_j\Gamma_{j'}}\over 2(\omega+i\Lambda)},~~{\rm where}~~
\Gamma_j=2\pi\Omega_j^2\varrho \,.
\end{align}
Substituting this result into Eq.~(\ref{a2}) and solving this algebraic equation, we find the amplitudes $\tilde b_{1,2}(\omega )$. The time-dependent amplitudes are obtained via the inverse Laplace transform,
$b_{1,2}(t)=\int_{-\infty}^\infty \tilde b_{1,2}(\omega )e^{-i\omega t}d\omega /2\pi$.

Consider, for simplicity, the case of aligned  levels $E_1=E_2=E$ and $\Gamma_1=\Gamma_2=\Gamma$. We find
$\{b_1(t),b_2(t)\}^T= U(t)\{b_1(0),b_2(0)\}^T$,
where
\begin{align}
U(t)={1\over2}\left(
\begin{array}{cc}
a(t)+1&a(t)-1 \\[5pt]
a(t)-1&a(t)+1
\end{array}
\right) \,,
\label{proj}
\end{align}
and
\begin{align}
a(t)={1\over A_+^{}-A_-^{}}(A_+^{}e^{-A_-^{}t}-A_-^{}e^{-A_+^{}t})
\label{proj0}
\end{align}
with $A_{\pm}^{}=[\Lambda -iE\pm \sqrt{(\Lambda -iE)^2-4\Gamma\Lambda}]/2$.

The null-result measurement in the reservoir, quantum mechanically, collapses the entire wave function  onto subset of the dot's states $|\Psi(t)\rangle\to
U(t)|\Psi(0)\rangle/N$, where $N$ is a normalization factor. After $n$ such null-result measurements in the reservoir with the subsequent time interval $\tau=t/n$, the final state of the system is $\{b_1^{(n)}(t),b_2^{(n)}(t)\}^T= U^n(\tau)\{b_1(0),b_2(0)\}^T/{\cal N}_n$, where ${\cal N}_n=[|b_1^{(n)}(t)|^2+|b_2^{(n)}(t)|^2]^{1/2}$ is a normalization of factor. One easily finds from Eq.~(\ref{proj})
\begin{align}
U^n_{}(\tau)={1\over2}\left(
\begin{array}{cc}
a^n_{}(\tau )+1&a^n_{}(\tau )-1 \\[5pt]
a^n_{}(\tau )-1&a^n_{}(\tau )+1
\end{array}
\right) \,.
\label{proj1}
\end{align}

Let us consider the limit of continuous measurement $n\to\infty$ by taking the frequency of measurements $\nu=1/\tau\to \infty$. In the same way we increase the bandwidth $\Lambda$ so that the ratio $x=\Lambda /\nu$ remains constant. Now we demonstrate that the final state becomes a function of the variable $x$ only. Consider the case of $E=c\Lambda$ so that the dot's level is always inside the band ($c< 1$) or outside it ($c>1$) with an increase of $\Lambda$. Alternatively, we can set the dots level $E=E_0$ so that it is inside the band when $\Lambda\gg E_0$. This essentially corresponds to the previous case with $c=0$. One finds from Eq.~(\ref{proj0}) that $A_+^{}= \kappa\Lambda -\Gamma/\kappa$ and $A_-^{}=\Gamma/\kappa$ [up to the order of $(\Gamma/\Lambda )^2$], where $\kappa=1-ic$. Then
\begin{align}
a^n_{}(\tau)=\left({1-{\Gamma\over\kappa^2\Lambda}\over 1-{2\Gamma\over\kappa^2\Lambda }}\right)^ne^{-{\Gamma t\over\kappa}}\left[1-{{\Gamma\over\Lambda}\over \kappa^2-{\Gamma\over\Lambda}}e^{-\big(\kappa -{2\Gamma\over\kappa\Lambda}\big)x}\right]^n
\nonumber
\end{align}
where $n=t\nu =t\Lambda /x$. Using $\big(1-{z\over n}\big)^n=e^{-z\big(1+{z\over 2n}+\cdots\big)}$, one obtains
\begin{align}
\left(1 -{\Gamma n\over\kappa^2\Lambda n}\over 1 -{2\Gamma n\over\kappa^2 \Lambda n}\right)^n=\left(1 -{\Gamma t \over\kappa^2x n}\over 1 -{2\Gamma t\over\kappa^2 xn}\right)^n=e^{{\Gamma t\over\kappa^2 x}\big(1+{3\Gamma\over 2\kappa^2\Lambda}\big)} \,.
\label{rengam}
\end{align}
Neglecting small terms $\sim \Gamma/\Lambda$ in the exponent of Eq.~(\ref{rengam}), we arrive at
\begin{align}
\bar a(t)=a^n_{}(\tau)=\exp\left({\Gamma t\over \kappa^2x}(1-e^{-\kappa x})-{\Gamma t\over\kappa}\right) \,.
\label{scal}
\end{align}
In the case of $E=E_0$ the result is the same, corresponding to $\kappa =1$. The probability of finding the electron in the upper dot (Fig.~\ref{fig1}), subjected to the null-result monitoring of the reservoir, is
\begin{align}
P_1^{}(t)={[\bar a(t)+1]^2|b_1(0)|^2+[\bar a(t)-1]^2|b_2(0)|^2\over 2[ a^2_{}(t)+1]} \,.
\label{nmeasur}
\end{align}

Equations~(\ref{scal}) and (\ref{nmeasur}) represent the evolution of a quantum system under continuous measurement. They show an explicit scaling in the $x=\Lambda \tau$ variable. It is remarkable that these equations can display either no influence of measurement or the Zeno effect, depending on the value of $x$. Indeed, in the limit $x\to\infty$ one finds $\bar a(t)\to e^{-(\Gamma/\kappa) t}$. The same is obtained from Eq.~({\ref{proj0}), presenting the Schr\"odinger evolution without intermediate measurements. On the other hand, in the limit $x\to 0$, one finds from Eq.~(\ref{scal}) that $\bar a(t)=1$, so $P_1(t)=1$, which corresponds to freezing of the electron in its initial state. This tells us that the variable $x$ must replace the measurement time $\tau$ for a description of the continuous measurement.

In Fig.~\ref{fg2} we plot the conditional probability $P_1(t)$ of finding the electron at time $t$ in the initial state, corresponding to the occupied upper dot (Fig.~\ref{fig1}). The solid lines in Figs. 2(a) and 2(c) correspond to Eq.~(\ref{nmeasur}) for $b_1(0)=1$ and $b_2(0)=0$ for different values of $x$. The dots correspond to numerical evaluations of $|b_1^{(n)}(t)|^2$ using Eqs.~(\ref{proj})$-$(\ref{proj1}) for $\Lambda=3\Gamma$ and, respectively, $\tau=x/\Lambda$ and $n=t/\tau$. One finds that simple analytical formulas, (\ref{scal}) and (\ref{nmeasur}), derived in the limit $\Gamma/\Lambda\ll 1$ can be applied with high accuracy even outside this limit, when $\Lambda$ is relatively low.  The same excellent agreement of Eqs.~(\ref{scal}) and (\ref{nmeasur}) with the numerical calculation is obtained for  $E_1=E_2=\Gamma$. The results that are not presented here, are almost  identical to those shown in Fig. 2(a) for $E_1=E_2=0$.

Figure 2(b) shows the results for slightly misaligned levels $E_{1,2}=\pm 0.05\Gamma$.
Unfortunately, for this case we did not find any simple analytical expression $\bar a(t)$ similar to  Eq.~(\ref{scal}). Therefore, instead of solid lines as in Figs. 2(a) and 2(c), we show by dashed lines numerical results, corresponding to $\Lambda =20\Gamma$. One can see that these lines coincide with the dots obtained from numerical calculations with $\Lambda =3\Gamma$, which again confirms the scaling in the $x=\Lambda/\nu$ variable even for non aligned levels. Note a quite different time dependence of $P_1(t)$ in a comparison with the aligned levels [Fig. 2(a)] for $x=0.2$ and  $0.02$ (strong non-Markovian case). However, for $x=2$, the effect of the levels misalignment in $P_1(t)$ is small. This can be anticipated since with an increase of $x$ we approach the Markovian limit, where the conditional probabilities for the aligned and misaligned dot levels differ very little \cite{SG11a}. Thus, for large $x$ Eqs.~(\ref{scal}) and (\ref{nmeasur}) can be still applied for slightly misaligned levels.

\begin{figure}[h]
\center
\includegraphics[scale=0.62]{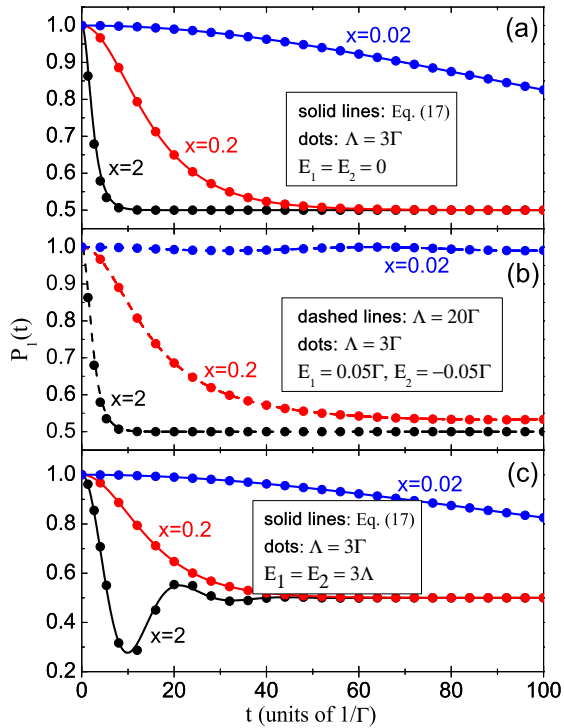}
\caption{(Color online) Occupation probability of the upper dot in Fig.~\ref{fig1}, conditioned on the null-result monitoring of the reservoir for different $x$ and bandwidths $\Lambda$: (a) $E_1=E_2=0$, (b) $E_1=0.05 \Gamma$ and $E_2=-0.05\Gamma$,
and (c) $E_1=E_2=E=3\,\Lambda$.
}
\label{fg2}
\end{figure}

The general behavior of the conditional survival probability and in particular the scaling in the $x$ variable, displayed in Fig.~\ref{fg2}, is in full agreement with a simple explanation of the Zeno effect, based on the energy-time uncertainty relation.  Indeed, the probability of transmission to the reservoir (Fig.~\ref{fig1}), is maximal when the dot's levels are in resonance with the peak of the density of states ($E_r=0$). If the reservoir is monitored with frequency $\nu$ [Fig.~\ref{fig1}(a)] its energy spectrum is shifted up or down by $\Delta E=\nu$. Equivalently, for the setup, shown in Fig.~\ref{fig1}(b), the energy levels of both dots are simultaneously shifted by the same energy $\nu$, whereas the reservoir spectrum remains unshifted by the measurements. As a result, the energy levels of the quantum dots are off-resonance with the density-of-state peak, so the probability of the interdot transmission through the reservoir decreases. In the limit $\nu\to\infty$, the energy shift becomes so large that the corresponding density of states is zero. As a result, the electron remains locked in its initial state.

For Markovian reservoirs, however, the density of states is constant ($\Lambda\to\infty$). Therefore, the shift of the reservoir spectrum [or the dots levels in Fig.~\ref{fig1}(b)] by the measurement is irrelevant for the transition rates. As a result, we expect no Zeno effect at all \cite{eg}. This does not contradict its derivation [Eqs.~(\ref{a10})-(\ref{a14})] based on the assumption that the evolution operator can be expanded in powers of $\tau$. Indeed, this assumption is not valid for Markovian reservoirs since the evolution operator is singular at $\tau=0$. It appears, for instance, in the divergence of the coefficient $C$ in (\ref{a13}) for the Markovian case.


We therefore demonstrated that the Zeno effect has nothing paradoxical in its nature by explaining it through the energy-time uncertainty relation.
Indeed, the large energy transfer does not necessarily destabilize the system due to the short-time measurements. This could happen only if there exist available reservoir states with such large energies. Otherwise the system cannot move since any quantum transitions between states with largely different energies are strongly suppressed. For instance, it can take place for reservoirs with a finite bandwith [Eq.~(\ref{lor})]. For Markovian reservoirs, however, the Zeno effect is not expected.

It is quite remarkable that precisely the {\em absence} of the Zeno effect can result in the paradoxical behavior of a quantum system under continuous measurement. Indeed, consider again the setup in Fig.~\ref{fig1}. Note that the two dots are connected only through the reservoir [Eq.~(\ref{a1})]. Nevertheless, for the Markovian case ($\Lambda\to\infty$), the electron can make transitions between the two dots without any record in the reservoir, even though the latter is continuously monitored. This appears to be a teleportation phenomenon   in its literal meaning, namely, undetectable  matter transfer between two distant places \cite{note2}. A similar phenomenon was discussed earlier for different systems, but without continuous monitoring \cite{fu}.

An experimental realization using the PC detector, shown in Fig.~\ref{fig1}(b), looks very promising. This detector is proven very efficient for single-electron monitoring \cite{michal}.
The measurement time can be varied by increasing the signal $\Delta I$ through an increase of the voltage. Alternatively, one can use another measurement device, for instance, a single electron transistor \cite{devoret}. In any case, a simultaneous monitoring of two dots in this type of experiment seems easier than continuous monitoring of the reservoir.

In conclusion, we have presented a quantum-mechanical analysis of electron transfer through a non-Markovian reservoir under continuous null-result monitoring. We found that the results differ from predictions of the QT method, except in the Markovian case. This suggests that the quantum trajectory method should be modified for non-Markovian environments by including explicitly the measurement time. The latter should appear in combination with the reservoir bandwidth.
We believe that our Eq.~(\ref{scal}),
which displays the scaling in the $\Lambda/\nu$ variable
and covers the Markovian and non-Markovian cases at once,
could be very useful for a possible extension of the QT approach to the non-Markovian case.
We also proposed a simple explanation of the Zeno effect without explicit use of the projection postulate. Finally, we discussed the undetectable quantum transfer through a continuum and its relation to the Zeno effect.

\begin{acknowledgements}
S.G. acknowledges the Beijing Normal University
and the Beijing Computational Science Research Center,
for supporting his visit.
This work was supported by the Israel Science Foundation
under Grant No.\ 711091, the NNSF of China under
Grants  No.\ 101202101 and No.\ 10874176, and the Major State Basic Research Project of China under
Grants No.\ 2011CB808502 and No.\ 2012CB932704.
\end{acknowledgements}


\begin{references}
\bibitem{note1} In fact, the ratio $\gamma=\Omega_{1r}/\Omega_{2r}$
    will oscillate with a frequency $\sim E_r^{1/2}l$, where $l$ is the distance between two matching points of the quantum dots with the reservoir. Therefore, in order to make $\kappa$ energy independent, two quantum dots are coupled to the reservoir (quantum wire) at the same end, as shown in Fig.~\ref{fig1}.
\bibitem{note3} If the levels  are not exactly aligned $E_1\not =E_2$, the state $|1'\rangle$ eventually decays to the reservoir. However, if the misalignment is smaller than the level width, the electron would occupy the upper dot for a long time and therefore can be observed there with large probability \cite{SG11a}.
\end{references}
\end{document}